\documentclass[prb,twocolumn,showpacs,preprintnumbers,amsmath,amssymb]{revtex4}
\usepackage{graphicx}

\begin{document}

\title{Density-Functional-Based Determination of Vibrational Polarizabilities 
in Molecules}
\author{Mark R. Pederson$^{1}$}\email{pederson@dave.nrl.navy.mil}
\author{Tunna Baruah$^{1,2}$}
\affiliation{
$^1$Center for Computational Materials Science, Code 6390,
Naval Research Laboratory, Washington DC 20375 \\
$^2$Department of Electrical and Computer Engineering, Howard University, 
Washington DC 20059}
\author{Philip B. Allen$^3$ and Christian Schmidt$^3$}
\affiliation{$^3$Department of Physics and Astronomy\\
State University of New York, Stony Brook, NY 11794-3800}
\date{\today}

\begin{abstract}
We develop a direct derivation for the primary contribution to the
vibrational polarizability for molecules, clusters and other finite
systems. The vibrational polarizability is then calculated within
the generalized gradient approximation to the density functional
theory for a variety of molecules and clusters. The agreement between
theory and experiment is quite good. The results show that for small
ionic molecules and clusters, inclusion of the vibrational polarizability
is necessary to achieve quantitative accuracy.
\end{abstract}

\maketitle

\maketitle


\section{\label{sec:int}Introduction}


Materials with high dielectric constants have many important technological
applications. For molecular assembled materials, the Clausius-Mossotti relation 
tells us that the dielectric constant is related in a simple way to the 
molecular polarizability and points to critical densities at which this
relation is expected to diverge. Predicting the {\it total} second-order molecular 
polarizability is one capability that is required for the computational design of 
materials with high dielectric constants.
The second largest contribution to the static second-order polarizability
tensor is generally accepted to be due to field-induced atomic relaxation. 
As the effect is of interest to several fields of research, a common terminology
is lacking. This effect has been referred to as the displacement-, atomic-, 
nuclear-, relaxation- or vibrational polarizability. Here we 
adopt the latter term and determine this effect within the double harmonic
approximation.~\cite{bishkirt}  As discussed below, this effect is governed by the
dynamic effective charge tensor which is known to account for
infrared intensities of vibrational modes in molecules and clusters.

The dynamic effective charge tensor describes how the
total dipole moment of a molecule or other finite system
changes due to an atomic displacement.  For a simple dipole consisting
of two point charges $\pm Q$, the change of the dipole
per unit change of the separation is just Q.
For crystals, the local dipole is not necessarily a well-defined notion,
and effective charge tensors come in multiple 
forms \cite{Gonze}. However, the lowest nonzero moment of a 
finite system is a well defined quantity. For a 
neutral molecule with $N$ atoms, both the             
electrical dipole moment $\vec{p}$ and the derivative of the dipole
moment with respect to  the $i$'th atomic position $\vec{u}_i$ are 
well defined. Only the latter quantity
is uniquely defined for a charged molecule.  These derivatives
may be expressed as a $3\times 3N$ tensor ${\sf Z}$ which
has units of charge, and is written
\begin{equation}
Z_{\alpha,i\mu}=\partial p_{\alpha} / \partial u_{i\mu}.
\label{eq:Z}
\end{equation}
This tensor is also sometimes called the ``polar tensor''
\cite{Morcillo}, and it is used for the calculation of the
infrared intensity \cite{Morcillo, Person, Porezag}.  
The infrared intensities are also
related to the vibrational component of the {\it dc} molecular 
polarizability~\cite{Gussoni}, and a simple proof
of this is included below.

By vibrational polarizability, we refer to the following physics. 
When a molecule is placed in a static electric field, it can lower
its energy through several mechanisms. First, the electronic clouds
rearrange themselves in response to the field which leads to an
induced electronic dipole moment given by $p_{{\rm el},x} 
= \Sigma_y \alpha_{{\rm el},xy} E_y$. This is
generally the largest linear effect. Second, this induced dipole moment
is further modified since the atomic positions rearrange
themselves in response to the forces caused by the direct application of
a field and the subsequent electronic rearrangement.                       
The tensor that describes the portion of the induced dipole
moment due to atomic rearrangement is what we refer to as the vibrational
polarizability. 

To concentrate on effects due to vibrational polarizability, we 
ignore molecular rotation and assume the molecule to be oriented in the 
lab frame.  Equivalently, we work in a frame which is tied to the molecule, 
so that there is a fixed dipole moment.  The polarizability
${\bf \alpha}$ is then a $3\times 3$ tensor which reduces to
a scalar for symmetrical molecules such as CH$_4$ or SF$_6$.
The molecular vibrations within the harmonic approximation
correspond to the classical normal modes of a coupled system of
oscillators
\begin{equation}
M_i \frac{d^2 u_{i\mu}}{dt^2}=-\sum_{j\nu}K_{i\mu,j\nu}u_{j\nu},
\label{eq:Newton}
\end{equation}
where the $3N\times 3N$ force constant tensor ${\sf K}$
is defined as
\begin{equation}
K_{i\mu,j\nu}=\frac{\partial^2 {\cal E}}{\partial u_{i\mu}
\partial u_{j,\nu}}
\label{eq:K}
\end{equation}
and ${\cal E}$ is the total energy of the molecule at zero field.

The dipole moment $\vec{p}$ is a first derivative of
the energy (${\cal E}$) and the dynamical charge tensor and the
electronic polarizability tensor are second partial derivatives 
given by:
\begin{equation}
p_{\alpha}=-\partial {\cal E}/\partial E_{\alpha},
\label{eq:mu}
\end{equation}
\begin{equation}
Z_{\alpha,i\mu}=-\partial^2 {\cal E}/\partial E_{\alpha}
\partial u_{i\mu} = \partial { F_{i\mu}}/\partial E_{\alpha},
\label{eq:Z2}
\end{equation}
\begin{equation}
\alpha_{{\rm el},\alpha\beta}=-\partial^2 {\cal E}/\partial E_{\alpha}
\partial E_{\beta}.
\label{eq:alpha}
\end{equation}
In Eqs.~(4-6), the electronic degrees of freedom
must be relaxed in response to changes of the independent variables 
(${\vec E},{\vec u_{1}},...,{\vec u_{N}})$ and the derivatives are
evaluated at zero field and displacement $({\vec E}={\vec u_i}=0)$.
Eq.~\ref{eq:Z2} also shows that the dynamical charge tensor determines
how the Hellmann-Feynman (HF) force $({ F_{i\mu}}=-\partial {\cal E}/\partial u_{i\mu})$ 
changes due to the application
of an external electric field. As discussed in Ref.~\onlinecite{Porezag}, 
the relationship
between the dynamic effective charge tensor and the derivative of
the HF force is both instructive and optimal
for efficient determination of infrared and Raman
intensities.

Now the total energy of the molecule may be expanded as a Taylor
series in powers of both the atomic displacements and applied
electric fields according to
\begin{equation}
{\cal E}={\cal E}_0 -\vec{p}\cdot\vec{E}
-\frac{1}{2} \vec{E}\cdot {\bf \alpha_{el}}\cdot\vec{E}
-\vec{E}\cdot {\sf Z}\cdot {\bf u}
+\frac{1}{2} {\bf u}\cdot {\sf K} \cdot {\bf u}.
\label{eq:Eexp}
\end{equation}
In the above equation $\vec{p}$ and $\alpha$ are the zero-field
values of the dipole moment and polarizability respectively.
The tensor notation is fairly obvious except perhaps for
the asymmetrical tensor ${\sf Z}$ whose transpose ${\sf Z}^{\rm T}$
is defined by
\begin{equation}
\vec{E}\cdot {\sf Z}\cdot {\bf u}={\bf u}\cdot
{\sf Z}^{\rm T}\cdot\vec{E}.
\label{eq:transpose}
\end{equation}

Now if a static external field $\vec{E}$ is applied, the
atomic coordinates ${\bf u}$ will relax to new positions
to  minimize the energy according to
\begin{equation}
{\bf u}={\sf K}^{-1}\cdot{\sf Z}^{\rm T}\cdot\vec{E},
\label{eq:neweq}
\end{equation}
and the corresponding energy of the relaxed molecule is
\begin{equation}
{\cal E}={\cal E}_0 -\vec{p}\cdot\vec{E}
-\frac{1}{2} \vec{E}\cdot [{\bf \alpha_{\rm el}}+{\bf \alpha_{\rm vib}}].
\cdot\vec{E}
\label{eq:newen}
\end{equation}
In the above, the vibrational part of the polarizability is given by
\begin{equation}
{\bf \alpha_{\rm vib}}={\sf Z}\cdot{\sf K}^{-1} \cdot{\sf Z}^{\rm T}.
\label{eq:alphavib}
\end{equation}
In fully indexed Cartesian form, the polarizability matrix is
\begin{equation}
\alpha_{{\rm vib},\alpha\beta}=\sum_{i\mu,j\nu}
Z_{\alpha,i\mu}(K^{-1})_{i\mu,j\nu}Z^{\rm T}_{j\nu,\beta}.
\label{eq:alphavib1}
\end{equation}
While the above expression clearly exhibits the isotopic independence
of this part of the polarizability tensor, a
simpler expression, directly comparable to experimental observables, 
is possible by rewriting this energy in terms of the normal modes of vibration.  Let
$|v>$ denote the eigenvector and $\omega_{v}$
the corresponding eigenfrequency, which satisfies the
Newtonian equations
\begin{equation}
{\sf K}|v>=\omega_v^2 {\sf M}|v>,
\label{eq:Newton1}
\end{equation}
where ${\sf M}$ is the mass tensor which in the atom displacement
basis $(i\mu)$ is
\begin{equation}
M_{i\mu,j\nu}=M_i \delta_{ij}\delta_{\mu\nu}.
\label{eq:M}
\end{equation}
The orthogonality and completeness relations are
\begin{equation}
<v|{\sf M}|v^{\prime}>=\delta_{v^{}v^{\prime}}
\end{equation}
\begin{equation}
\sum_v |v><v|={\sf M}^{-1}
\end{equation}
The force constant matrix can be written as
\begin{equation}
{\sf K}=\sum_v {\sf M}|v>\omega_v^2 <v|{\sf M}
\end{equation}
\begin{equation}
{\sf K}^{-1}=\sum_v |v>\omega_v^{-2} <v|.
\end{equation}
The effective charge tensor can now be written in
the eigenvector basis as the charge vector for each
normal mode
\begin{equation}
Z_{\beta,v}=\hat{\beta}\cdot{\sf Z}|v>.
\end{equation}
Then the vibrational polarizability can be written
as a sum of contributions from the normal modes,
\begin{equation}
{\bf \alpha}_{{\rm vib},\alpha\beta}=\sum_v Z_{\alpha,v}
\omega_v^{-2} Z^{\rm T}_{v,\beta}.
\end{equation}
This equation is a generalization of a known relation \cite{Gussoni}
between infrared intensities and static polarizability.  In the past, this
equation has been used to determine vibrational polarizabilities from 
experimental IR data and from calculations.~\cite{ref1} We include our derivation
here because it appears to be rather simple in  comparison to previous derivations 
that appear in the literature. Eq.~(20) follows immediately from Eq.~(17.29) of
Born and Huang.~\cite{born-huang} It has also been derived 
by Flytzanis.~\cite{flytzanis}
Probably the earliest modern discussion of vibrational polarizabilities 
using quantum-mechanical derivations can be found in Ref.~\onlinecite{bcb,dblc}
where applications to CHCl$_3$ and CHF$_3$ are discussed and the above
formula is derived within a sum over states method within the clamped
nucleus approximation. Eq. (1) and A5 of Ref.~\onlinecite{bcb}b lead to
our Eq.~(20).  However, as noted
in Refs~\onlinecite{bcb} and \onlinecite{dblc}, one of the earliest discussions dates back
to 1924.~\cite{1924}

In addition to the interaction discussed above, there are other smaller vibrational 
effects that modify the polarizability of a molecule. The presence of the field modifies
the spring constant matrix which changes the zero-point energy of the molecule. Also,
the occurrence of anharmonicity, both diagonal and off-diagonal, leads to further 
corrections. We are unaware of discussions on the role of 
off-diagonal anharmonicity, but 
discussion of the zero-point effect and diagonal anharmonicity may be found
in Ref.~\onlinecite{Marti}.  In the notation of the work of Marti and Bishop,
the above term is equivalent to $[\mu^2]^{0,0}$ in their paper.

\section{Computational Details}
The calculations presented below have been performed using the
NRLMOL suite of density-functional-based cluster codes.~\cite{nrlmol}
The Perdew-Burke-Ernzerhof (PBE) energy functional has been used
in all calculations.~\cite{gga}  The Kohn-Sham equations are solved 
self-consistently for 
each electron in the problem. Then
the HF forces are 
calculated and the geometries are updated using standard force optimization
methods. Geometries were
considered converged when the force on each
atom fell below 0.001 Hartree/Bohr. However, for the Na and H$_2$O 
clusters we used a tighter
force convergence criteria of 0.0001 Hartree/Bohr. The numerical integration
mesh was also significantly more dense for our calculations on the water
molecules.
The method for generating the 
basis sets used for these calculations is discussed in 
Ref.~\onlinecite{Porezag-bas}. Basis sets and the unpublished geometries are  
available upon request. Once the optimized geometries~\cite{bp,bzrp,bplc} 
are obtained, the vibrational frequencies, eigenvectors, and dynamical charge 
tensors
(Z$_{\alpha,i \mu}$) are determined using the method discussed 
in Ref.~\onlinecite{Porezag}. We then use Eqn. (20) to determine the vibrational
component of the polarizability.  As discussed in Ref.~\onlinecite{Porezag}, the
infrared and Raman spectra showed some sensitivity to the inclusion of
longer range polarization functions. We have used such polarization
functions for the calculations displayed in Table. I.
\section{Results}

Table I presents calculations on several molecular systems
which include both covalent, ionic bonding and hydrogen bonding. It also
include calculations on systems with both loosely and tightly bound
electrons.  We have calculated both the electronic and vibrational contributions
to the polarizability tensor. Agreement is generally good.              

{\bf Fullerene Molecule:} The polarizability of the fullerene molecules has
been well studied both theoretically and 
experimentally.~\cite{quong,ecklund,bonin,antoine,ballard} Here we calculate
the electronic polarizability to be 82.9 $\AA^3$ which is in good agreement
with one of the earliest density-functional calculations~\cite{quong} of 
83.5 $\AA^3$. This earlier calculation used the same code, a 
slightly different version of DFT, slightly smaller basis sets and 
geometries that were not as well
converged.~\cite{quong} The good agreement between the early and most 
recent calculations indicate that the electronic part of the neutral 
fullerene polarizability is rather robust, and the experimental 
polarizability~\cite{ecklund,bonin,antoine,ballard}
is known to be very close to this number as shown in Table I. 
Based on experiments, it has also been suggested that the
polarizability due to lattice relaxation is 2 $\AA^3~$~\cite{ecklund} 
which is small
but still four times larger than the value calculated here. The deviation
may be due to the lower T$_h$ symmetry that occurs when the icosahedral 
C$_{60}$ molecules are placed on a cubic lattice. Such a symmetry lowering
would cause some of the optically silent G$_u$ and H$_u$ modes to split and 
partially fall into the IR active T$_{1u}$ manifold which in turn could
lead to additional vibrational polarizability.  There will also be weak
IR activity due to weak intermolecular vibrations activated by weakly 
broken translational symmetry.~\cite{extrair} 
Also included in Table I are the electronic and
vibrational polarizabilities of a C$_{60}$ molecule with an endohedral 
Kr atom. The addition of the Kr atom adds another infrared mode due to a
rattling motion of Kr inside the C$_{60}$ cage.  The low frequency Kr
rattling mode is found to be at 88 cm$^{-1}$  but the IR intensity associated
with this mode is 1000 times smaller than the four T$_{1u}$ modes associated
with the fullerene cage. Because of this the vibrational polarizabilities
are unchanged due to the addition of an inert endohedral atom.

{\bf Acetylene:} The acetylene molecule provides an interesting test case 
because the anisotropy of the polarizability tensor is reversed significantly
by the inclusion of the vibrational terms. For example, in $\AA^3~$ the 
electronic and vibrational polarizability tensors have been measured to be 
(2.43, 2.43, 5.12) and (0.667, 0.667, 0.027), respectively. Density functional
theory yields (2.96, 2.96, 4.78)  and  (0.71, 0.71, 0.030)  $\AA^3~$
which is in reasonably good agreement with experiment.

{\bf Halogen containing Ionic Molecules:} 
Halogen containing compounds are known
to exhibit high vibrational polarizabilities as would be expected since they
make very good ionic systems.~\cite{bcb,dblc}
We have performed calculations on NaF, SiF$_6$,
SiF$_4$ and TiCl$_4$. Of the molecules in this size regime listed in the 
large database of Gussoni, the latter three stand out as having very large
vibrational contributions. The agreement between theory and experiment is
in the neighborhood of 15 percent for these systems.

{\bf Isomeric dependencies:} Acetonitrile (CH$_3$CN) and methylisonitrile
(CH$_3$NC)
have the same chemical composition. However, the former has the two
carbon atoms bound to one another while the latter has the nitrogen bound to
the methyl radical.  This causes a five percent difference in the 
electronic polarizability and a factor of two difference in the vibrational
polarizability. The source of the deviation in the vibrational polarizability
is clearly due to changes in the spring constant matrix since Eq.~(11) shows
that changes of mass cannot perturb the vibrational contributions within the double 
harmonic approximation.~\cite{bishkirt}  There is a one-fold mode at 2269 1/cm for acetonitrile
that is reduced to 2149 1/cm for methylisonitrile. In addition to 
a reduction in the vibrational frequency, the infrared intensity of the 
methylisonitrile is 2.57 compared to 0.227 in the case of acetonitrile. 
This mode accounts for about 75 percent of the difference in the vibrational 
polarizability. The large change in infrared intensity in this frequency
range should be a clear indicator of methylisonitrile isomerization to
acetonitrile at higher temperatures.

{\bf Sodium clusters:} In two recent papers~\cite{zope} Blundell, Guet and Zope 
and Kronik, Vasiliev and Chelikowsky have calculated the
temperature dependence of polarizabilities in sodium clusters. They show that
temperature effects enhance the apparent polarizability at 300K. 
This temperature enhancement appears to account for most of the difference
between experiment and the calculated electronic polarizabilities from many 
different theoretical 
calculations.\cite{moullet,Guan,Rubio,Pacheco,Vasiliev,calaminici} 
Our results show that the vibrational
contribution to the polarizability is indeed small for the sodium clusters
which supports the assertion that temperature effects are important in 
these systems.

{\bf Weakly Bound Molecules:} As mentioned in our discussion of fullerene
molecules the vibrational polarizability between two weakly bound molecules
could be enhanced if the weak intermolecular vibrations are IR active. 
As discussed in Ref.~\onlinecite{molphys} the water dimer represents an
extreme example of this case. As shown in Table I, we find the electronic 
polarizability of this molecule (3.19 $\AA^3$) to be approximately twice that 
of a water monomer. The electronic polarizabilities obtained for the 
water trimer and pentamer also show a linear scaling as a function of
the number of molecules.  This result is in good aggreement with the work of 
Maroulis {\em et al}~\cite{jcp-mar} (2.90 $\AA^3$) and Eckart {\em et al} 
(2.48 $\AA^3$).\cite{molphys} 
Maroulis {\em et al}~\cite{jcp-mar} have carefully studied the electronic
polarizability as a function of both basis set and level of correlation. The
uncertainties due to these effects are at most 12.5 per cent indicating that
large deviations from these values must be due to other effects.
Our calculated double-harmonic vibrational polarizability of 23.30 $\AA^3$ is 
indeed a factor of seven times larger than the electronic polarizability. 
Eckart {\em et al} find this term to be even larger (39.2 $\AA^3$) and 
further demonstrate that anharmonic corrections enhance the vibrational component 
of the dimer by an additional factor of three. The large vibrational enhancement
in polarizability in going from the monomer to dimer is indeed interesting. In
particular, the scaling of this term as a function of system size is impossible
to guess based upon
the results of the monomer and dimer. It is reasonable to expect that this large
result should be an upper limit since a dielectric medium that is coupled to an IR active
mode should counteract the IR activity and thus the vibrational polarizability. So the
presence of more water molecules should lead to a vibrational polarizability that is 
eventually sublinear in the total number of molecules. To partially address
this point we have performed additional calculations on the trimer and pentamer.
Our results show a decrease in the total vibrational polarizability in going from
the dimer to the trimer and a flattening of the total vibrational polarizability
for the pentamer. For the pentamer the ratio of the vibrational to electronic
polarizability has decreased significantly from seven for the dimer to slightly
less than two for the pentamer. 
Overall, these results show that weak intermolecular
vibrations can enhance the vibrational polarizability over what is determined
from intramolecular vibrations.

\section{Summary}
We have presented a straightforward derivation for the vibrational
polarizability of a molecule. We have used the generalized-gradient approximation 
to the density functional theory to evaluate both the electronic polarizability and 
this vibrational correction.   In accord with experiment, our results show that this
term can be important in smaller ionic molecules and weakly bound systems, but that 
it is smaller in covalent systems or where the frontier electrons are delocalized.

\begin{acknowledgments}
The work was supported in part by NSF grants NIRT-0304122, HRD-0317607, 
ONR and the HPCMO CHSSI program. We thank D.M. Bishop and J.L. Feldman
for helpful advice.
\end{acknowledgments}


\begin{table}[bh!]
\begin{center}
\caption{Calculated and experimental vibrational 
polarizabilities ($\AA^3$) for molecules. This is 
one third of the trace of the polarizability tensor.
Unless otherwise stated the experimental values are 
taken from Ref.~\onlinecite{Gussoni}. 
Experimental
data for C$_{60}$ is from Refs.~\onlinecite{ecklund,bonin,antoine,ballard} and
references therein.
For the pure sodium 
clusters, the experimental values are total polarizabilities (denoted
with a *)
which have been taken from Refs.~\onlinecite{knight,molof}.  }

 ~ \\
\begin{tabular}{|l|r|r|r|r|}
\hline
Molecule&\multicolumn{2}{|c|}{Vibrational}& \multicolumn{2}{|c|}{Electronic} \\
\hline
\hline
           & Theory & Expt. & Theory & Expt.\\
\hline
H$_2$O       & 0.04   & 0.04  & 1.57  & 1.45 \\
(H$_2$O)$_2$ &23.30   &       & 3.19  &      \\
(H$_2$O)$_3$ &13.50   &       & 4.82  &      \\
(H$_2$O)$_5$ &14.52   &       & 8.13  &      \\
NF$_3$       & 1.15   & 0.70  & 3.07  & 2.81 \\ 
HCCH         & 0.48   & 0.45  & 3.56  & 3.40 \\ 
CH$_4$       & 0.04   & 0.03  & 2.63  & 2.60 \\ 
TiCl$_4$     & 2.04   & $>$1.68 & 15.03 & 15.0 \\ 
SF$_6$       & 2.29   & 2.29  & 5.15  & 4.49 \\
SiF$_4$      & 2.09   & 1.75  & 3.72  & 3.32 \\
HCN          & 0.21   & 0.14  & 2.62  & 2.59 \\
H$_3$CNC     & 0.07   &  --   & 4.87  &      \\
H$_3$CCN     & 0.03   & 0.04  & 4.59  & 4.28 \\
C$_{60}@Kr^a$  & 0.55   &       &83.3  &     \\
C$_{60}$     & 0.58   & 2.0   &82.9  & 83  \\
Na$_2$       & 0.00   &  --   &35.91  &37.91[*]  \\ 
Na$_3$       & 1.72   &  --   &60.89  &69.8[*]  \\ 
Na$_8$       & 0.63   &  --   &116.2  &133.5[*]  \\ 
NaF          & 0.29   &       & 2.71  &  \\ 
Ti$_8$C$_{12}^b$& 3.72 &        &56.40  &  \\
As@Ni$_{12}$@As$_{20}^c$ &4.56   &         &140.86 &     \\
\hline
\hline
\end{tabular}
\end{center}
$^{a,b,c}$Ref.~\onlinecite{bp,bplc,bzrp} respectively \\
\end{table}

\begin{thebibliography}{99}
\bibitem{born-huang}Born and Huang note that when a molecule is viewed
as a collection of normal modes with a linear perturbation of the form 
$\Sigma_j a_j q_j$ that  the energy decreases as 
$\frac{1}{2}\Sigma_j \frac{a_j^2}{\omega_j^2}$. In an actual molecule with
an electric field the coupling constant $a_j$ would be determined from
the ${\bf E} \cdot \frac{d{\vec p}}{dq_j}$ which would also lead to 
Eq.~(20)  of this paper.
\end{thebibliography}
\end{document}